\documentclass[aps,twocolumn,floatfix,bibnotes]{revtex4}

\usepackage{dcolumn}
\usepackage{color}
\usepackage{amsmath}

\usepackage[dvips]{graphicx}

\newcommand\pictc[5]{\begin{figure}[t]
                        \centerline{
                        \includegraphics[width=#1\columnwidth]{#3}}
                    \protect\caption{\protect\label{fig:#4} #5}
                     \end{figure}            }

\newcommand\pict[4][1.]{\pictc{#1}{!tb}{#2}{#3}{#4}}

\newcommand\rpict[1]{\ref{fig:#1}}
\newcommand\leqt[1]{\protect\label{eq:#1}}
\newcommand\reqtn[1]{\ref{eq:#1}}
\newcommand\reqt[1]{(\reqtn{#1})}

\newcommand{\be}{\begin{equation}}
\newcommand{\ee}{\end{equation}}
\newcommand{\bea}{\begin{eqnarray}}
\newcommand{\eea}{\end{eqnarray}}

\newcounter{Fig}

\begin{document}

\begin{sloppy}

\title{Surface Bloch waves in metamaterial and metal-dielectric superlattices}

\author{Slobodan M. Vukovic$^{1,2}$, Ilya V. Shadrivov$^1$, and Yuri S. Kivshar$^1$}

\affiliation{$^1$Nonlinear Physics Center, Research School of Physics and Engineering,
Australian National University, Canberra ACT 0200, Australia\\
$^2$Institute of Physics, University of Belgrade, Zemun 11080, Serbia}

\begin{abstract}
We study the properties of electromagnetic Bloch waves in semi-infinite periodic structures created by alternating
metamaterial and dielectric layers. We derive and analyze the dispersion relations in the long-wavelength limit for both TE- and TM-polarized  surface Bloch modes, for magnetic metamaterials with negative refraction and metal-dielectric plasmonic superlattices. We reveal that in the subwavelength regime the bulk modes are characterized by {\em three different refractive indices} ("tri-refringence") while the surface modes can propagate parallel to the Bloch wavevector and along the interface between superlattice
and semi-infinite dielectric.
\end{abstract}

\maketitle

Subwavelength confinement and manipulation of light in nanoscale optical structures is of a great importance for applications in imaging, sensing, high-resolution lithography, all-optical signal processing, photonic funnels
and superlenses~\cite{Pendry:2000-3966:PRL,Fang:2005-534:SCI,Ozbay:2006-189:SCI,Shin:2006-73907:PRL,Jacob:2006-8247:OE,Smith:2006-391:JOSB}. However, the diffraction limit prevents the light confinement on the scales smaller than a half of the wavelength. Conventional techniques
for the resolution improvement are based on the use of high-index materials~\cite{Fang:2006-9203:OE}, but the further progress in this field requires a design of artificial materials, often termed as metamaterials, with higher refractive
indices~\cite{Karalis:2005-63901:PRL,Govyadinov:2006-155108:PRB}. Plasmonic
macrostructure composites that operate in the regime of coupled waveguide modes, have been
recently proposed for a variety of future imaging systems and beam steering~
\cite{Alu:2004-199:ITMT,Podolskiy:2005-201101:PRB,Wangberg:2006-498:JOSB,Antosiewicz:2006-3389:OE}. At the same time, the absence of natural magnetism at optical and infrared frequencies requires again the fabrication of nanostructured metamaterials in order to achieve required dispersion properties for negative refraction~
\cite{Shadrivov:2003-3820:APL,Zhang:2005-4922:OE,Zhang:2005-137404:PRL,Dolling:2006-892:SCI}. A typical width of a single layer in such metamaterials is of the order of 10~nm, that is much smaller than the free-space wavelength of optical or
infrared radiation. Thus, it is usually assumed that optical properties of
those multilayered composites can be described by an effective-medium
approximation~\cite{Alu:2006-571:JOSB}, although the typical field variation
length may be less than the free-space wavelength~\cite{Elser:2007-191109:APL}.
On the other hand, many studies of negative-index materials that use
the original design based on simultaneously negative dielectric permittivity and magnetic permeability
assume that the homogenization limit for metamaterials is fulfilled, either
for isotropic~\cite{Veselago:1967-517:UFN,Shelby:2001-77:SCI,Lindell:2001-129:MOTL} or
`indefinite'~\cite{Smith:2003-77405:PRL} models. In particular, the effective-medium
approximation has been used to study surface waves at the interfaces between metamaterial and surrounding media that may lead to the subdiffraction imaging. 

In this Letter, we study the electromagnetic waves propagating in
two-dimensional semi-infinite periodic structure created by alternating
layers of metamaterial and conventional dielectric and semi-infinite homogeneous dielectric placed normally to the interfaces within the multilayered media [see the geometry in Fig.~\rpict{structure}(a)]. A unit cell of this layered structure is formed by two layers: one with the thickness $a_{1}$, dielectric permittivity $\varepsilon_{1}$, and magnetic permeability $\mu_{1}$, and the other one, with the parameters $a_{2}$, $\varepsilon_{2}$, and $\mu _{2}$, respectively. The layered media has the period $d=a_{1}+a_{2}$.
Such a structure supports the propagation of surface waves in the direction parallel
to the Bloch wavevector $k_{B}||z$ in the layered medium. Surface waves that propagate
along the interface between the layered dielectric media and a semi-infinite dielectric cladding
placed parallel to the interfaces within the multilayered medium (i.e. with
the wave vector $k_{\perp }||x$ perpendicular to the Bloch wavevector) were
studied more than thirty years ago in Refs.~\cite{Yeh:1977-423:JOS,Yariv:1977-438:JOS}.
Although many authors have investigated subsequently the same geometry, the surface waves
that propagate parallel to the Bloch wavevector, have not been analyzed yet, to
the best of our knowledge. We term those waves as {\em surface Bloch waves}.

There exist surface modes supported by each individual interface within the
layered media, but the coupling (weak or strong) of these modes gives birth
to new collective modes that propagate in the same direction, i.e.
perpendicular to the Bloch wavevector. In the regime of strong coupling,
when the unit cell thickness $d$ is much smaller than the free-space
wavelength $\lambda$, the optical properties of a multilayered structure
can be described by effective permittivity~\cite{Alu:2006-571:JOSB} $\hat{\varepsilon}$ and effective permeability $\hat{\mu}$. Generally speaking, these quantities are diagonal tensors with two equal components in the directions perpendicular to the Bloch wavevector: $\hat{\varepsilon}={\rm diag}(\varepsilon _{\perp },\varepsilon _{\perp},\varepsilon _{||})$; $\hat{\mu}={\rm diag}(\mu_{\perp},\mu_{\perp },\mu_{||})$.

Assuming that the electric field of the s-polarized (TE) mode and the
magnetic field of the p-polarized (TM) mode in the form: $\sim A_{k_{B}}(z)\exp {(ik_{B}z+ik_{\perp}x-i\omega t)}{\bf {y}}$, the dispersion of the Bloch modes in the infinite layered media is described by the well-known equation for {\em bulk Bloch waves} that comes from the eigenvalue problem for two-layer transfer matrix~\cite{Yeh:1977-423:JOS,Yariv:1977-438:JOS}
\bea\leqt{BlochModes}
\cos {(q_{B}\ d)} = \cos {(k_{1}\ a_{1})}\cos {(k_{2}\ a_{2})} - \nonumber\\
\frac{1+\alpha _{s,p}^{2}}{2\alpha _{s,p}}\sin {(k_{1}\ a_{1})}\sin {(k_{2}\ a_{2})}
\eea
where $k_{1,2}=(\varepsilon _{1,2}\mu _{1,2}-q_{\perp}^{2})^{1/2}$.
\pict[1]{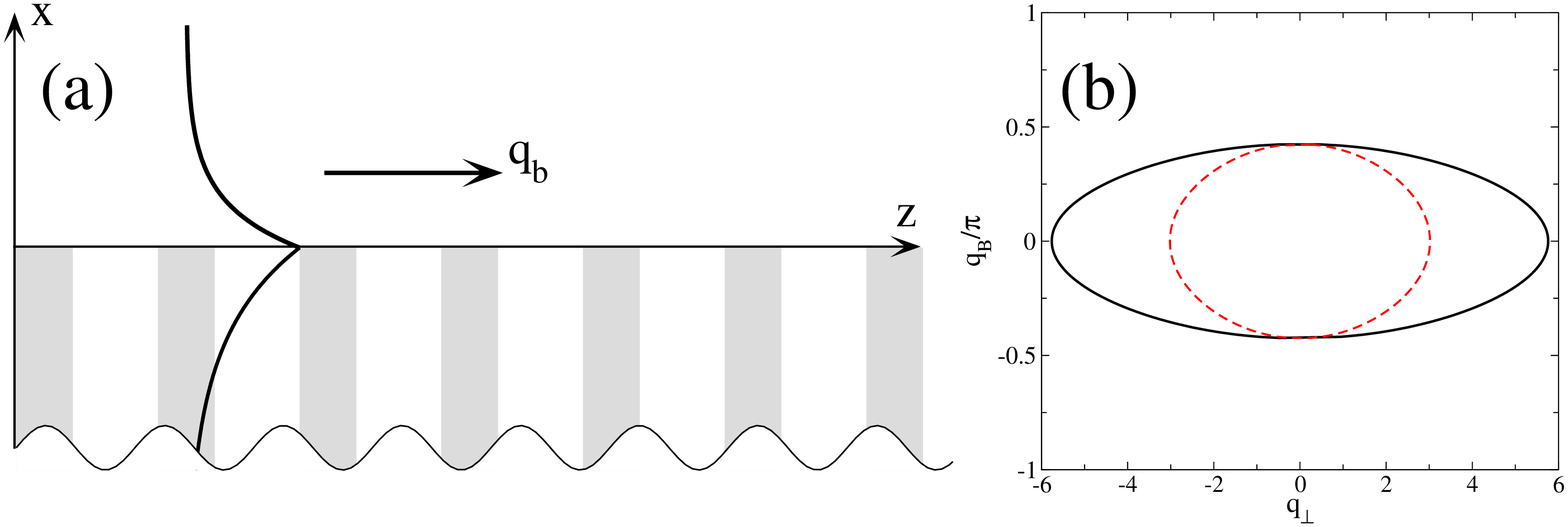}{structure}{(a) Geometry of the problem. (b) Equifrequency contours for
$\omega/\omega_p$ = 0.3 obtained from the dispersion (\ref{eq:TE_TM_Bulk}).}
We note that all spatial dimensions are normalized to $c/\omega$, and
wavenumbers to $\omega/c$. The polarization dependent
coefficients $\alpha_{s,p}$ are defined as $\alpha_{s}=k_{2}\mu
_{1}/k_{1}\mu_{2}$, and $\alpha_{p}=k_{2}\varepsilon_{1}/k_{1}\varepsilon_{2}$. In {\em the subwavelength} regime, when $|k_{1}a_{1}|\ll 1$, and $|k_{2}a_{2}|\ll 1$, after the Taylor expansion in Eq.~\reqt{BlochModes}, we
obtain the dispersion relations similar to those for the light propagation
in uniaxial crystals, for the s- and p-polarized waves
\be\leqt{TE_TM_Bulk}
\frac{\sin^{2}{q_{B}d/2}}{n_o^2} +
	\frac{q_{\perp }^{2}(d/2)^{2}}{\left(n_e^{s,p}\right)^2} =
	\frac{d^2}{4} ,
\ee
where $n_o = \sqrt{\varepsilon_{\perp }\,\mu_{\perp}}$ is the refractive index of the ordinary wave, $n_e^s = \sqrt{\varepsilon_{\perp}\, \mu_{\parallel}}$  and $n_e^p = \sqrt{\varepsilon_{\parallel}\, \mu_{\perp}}$ are the refractive indices of s- and p-polarized extraordinary waves, respectively. The effective material parameters appear as a result of the Taylor expansion
rather than an averaging procedure:
\bea\leqt{effective_params}
\varepsilon_{\perp} = \left(1-\delta \right) \varepsilon_1 +
	\delta \varepsilon_2; \ \ \
\varepsilon_{\parallel} = \frac{\varepsilon_1\,\varepsilon_2}{
	 \left(1-\delta \right) \varepsilon_2 +
	\delta \varepsilon_1}; \nonumber\\
\mu_{\perp} = \left(1-\delta \right) \mu_1 + \delta \mu_2; \ \ \
\mu_{\parallel} = \frac{\mu_1\,\mu_2}{
	\left(1-\delta \right) \mu_2 +
	\delta \mu_1},
\eea
where $\delta=a_2/d$, $0<\delta<1$ describes the degree of anisotropy. The structure becomes isotropic at $\delta=0$, and $\delta=1$, and anysotropic between this values.

Equation~(\ref{eq:TE_TM_Bulk}) describe the dispersion relations of bulk waves for infinite layered media in the long-wavelength approximation. For nonmagnetic media, and under the additional approximation $|q_{B}d|\ll 1$, these equations reduce to the equations obtained in Refs.~\cite{Yeh:1977-423:JOS,Yariv:1977-438:JOS},
where it was shown that a periodic stratified medium is formally analogous
to an uniaxial crystal, yielding for s- and p-polarizations, respectively:
$q_{B}^{2} + q_{\perp}^{2} = \varepsilon_{\perp}$, and $q_{B}^2/\varepsilon_{\perp} + q_{\perp}^2 / \varepsilon_{\parallel} = 1.$ Although these equations look very attractive (because they are formally analogous to the well known Fresnel equations),
it is worth noting that imposing $|q_{B}d|\ll 1$ leads to loosing zonal
structure and band gaps typical for Bloch waves. Under the same approximation, inclusion of higher order terms in the Taylor expansion revealed nonlocal effects (spatial dispersion) in nanolayered metamaterials~\cite{Elser:2007-191109:APL}. On the other hand,
inclusion of magnetic metamaterials gives rise to the appearance of the
extraordinary wave even in the case of s-polarization, if the formal analogy
with uniaxial crystals has been adopted. There are two ordinary s- and
p-polarized modes that obey the same dispersion law: $\sin ^{2}q_{B}d/2=\varepsilon _{\perp }\mu _{\perp }(d/2)^{2}$ and two extraordinary s- and p-polarized modes with different dispersions: $q_{\perp}^{2}=\varepsilon _{\perp }\mu _{||}$ and $q_{\perp }^{2}=\varepsilon_{||}\mu _{\perp }$, respectively. Thus, for the same frequency, we now have three different refractive indices, and when $|q_{B}d|\ll 1$, Eqs.~(\ref{eq:TE_TM_Bulk}) represent two ellipses in $(q_{B},q_{\perp })$-plane with one common axis (see Fig.~\rpict{structure} (b)), instead of a circle and an ellipse which we have for uniaxial crystals. So, magnetic metamaterials can produce either negative or normal (positive) refraction in all directions, as well as negative refraction in one direction and normal in the other one. We emphasize that for both s- and p-polarizations the
desired characteristics of the periodic stratified metamaterials can be
engineered by a variation of the thicknesses of the constitutive layers.

\pict{fig02}{params}{Effective dielectric and magnetic parameters of the metamaterial superlattices ($\delta = 0.95$).}

As an example, in Fig.~\rpict{params} we plot the effective parameters for
the case of a periodic structure with metamaterial layers in vacuum ($\varepsilon_1 = \mu_1 = 1$). We assume that the metamaterial layers have the frequency-dependent dielectric permittivity and magnetic permeability of the form: $\varepsilon_2 = 1 - \omega_p^2/\omega^2$ and $\mu_2 = 1 -
F\omega^2/(\omega^2 - \omega_g^2)$, where we take $F = 0.56$, $\omega_g^2/\omega_p^2 = 0.16$.

To study {\em surface Bloch waves} in long-wavelength limit, we assume that multilayered media is truncated, and it occupies a half-space $x<0$, while the other half-space ($x>0$) is occupied by homogeneous dielectric with dielectric constant $\varepsilon_c$. To obtain the solutions for surface waves, we consider in Eq.~(\ref{eq:TE_TM_Bulk})
purely imaginary values of $q_{\perp}$. Using the standard continuity
conditions of the tangential field components, we than obtain the dispersion
relation for the s-polarized surface waves,
\be
\leqt{s-disp}
\sqrt{\frac{\mu_{||}}{\mu_{\perp}} \left[ \sin^2 \left( q_B d/2\right) -  \varepsilon_{\perp} \mu_{\perp} (d/2)^2 \right]}= - \mu_{||} \frac{d}{2} \sqrt{q_B^2 - \varepsilon_c},
\ee
whereas for the p-polarized surface waves, we have
\be
\leqt{p-disp}
\varepsilon_c \sqrt{\frac{\varepsilon_{||}}{\varepsilon_{\perp}} \left[ \sin^2 \left( q_B d/2\right) -  \varepsilon_{\perp} \mu_{\perp} (d/2)^2 \right]} = - \varepsilon_{||} \frac{d}{2} \sqrt{q_B^2 - \varepsilon_c}.
\ee

Equations (\ref{eq:s-disp}, \ref{eq:p-disp}) are obtained in the approximation that the thicknesses of each layer is small, however we did not make any assumptions about the values for the Bloch wavenumber. However, since we used continuous medium approximation for matching boundary conditions on the interface, it is expected that the results are valid when the Bloch wavelength is larger than the period of the unit cell ($q_B d < 1$). More general case of arbitrary wavenumber will require numerical simulation in order to obtain dispersion properties of surface Bloch waves.

\pict{fig03}{disp_mm}{Dispersion of s- and p-polarized Bloch surface waves in metamaterial superlattices, for $\delta = 0.95$ and $d=0.1 c/\omega_p$. }

\pict{fig04}{disp_met}{Dispersion of p-polarized Bloch surface waves in metal-dielectric structures, for three different values of $\delta$ and $d=0.1 c/\omega_p$}

Now we calculate the dispersion of the surface Bloch modes with the effective parameters shown in Fig.~\rpict{params} for the waves propagating along the interface between the structured medium and vacuum ($\varepsilon_c = 1$). Dispersion of surface Bloch modes for metamaterial superlattices created by layers of metamaterial and conventional dielectric is calculated using Eqs.~(\ref{eq:s-disp},\ref{eq:p-disp}), and it is shown in Fig.~\rpict{disp_mm}. From the dispersion equations we see that the p-polarized waves can exist in the frequency ranges, where $\varepsilon_{\parallel} <0$, while s-polarized waves exist for $\mu_{\parallel} <0$. We found that for the values of the parameter $\delta$ close to $1$, there are in general two branches of p-polarized waves and one branch of s-polarized waves. Both upper and lower branches of the p-polarized waves are forward, while the s-polarized surface Bloch wave is backward (negative group velocity), and it exists in the range of frequencies $0.4 < \omega/\omega_p < 0.603$. When we decrease the parameter $\delta$  towards the value of $1/2$, the frequency range for the s-polarized wave existance becomes narrower, and it {\em becomes forward} (not shown), and then it disappears. P-polarized waves also disappear for values of $\delta<1/2$, and there are no surface waves for such values of $\delta$ within the limits of our approximation.

For nonmagnetic metamaterials or metals in the layered structure, we use the 
Drude-type formula for $\varepsilon _{2}(\omega )$, but consider $\mu_{2}=1$. We find no s-polarized surface waves, but for the p-polarized surface Bloch modes we obtain the dispersion shown in Fig. ~\rpict{disp_met}. Surface Bloch modes exist only when the metal occupies more than half of the structure, i.e., $1/2<\delta<1$. As we change the metal filling fraction from $\delta=1$, when the frequency range of the mode existence is $0<\omega/\omega_p < 1/\sqrt{2}$, to $\delta = 1/2$,  the frequency range of mode existence vanishes.

In conclusion, we have derived and analyzed the dispersion characteristics of the electromagnetic surface Bloch waves in metamaterial and metal-dielectric structures in the long-wavelength limit. We have revealed that in the subwavelength regime the bulk modes are characterized
by three different refractive indices while the surface modes can propagate parallel to the Bloch wavevector along an interface between two superlattices. As a special case, we have studied the propagation along an interface between the superlattice and a semi-infinite dielectric in truncated layered structures. We note that the short-wavelength limit requires intensive numerical simulations and it will be studied elsewhere.

The work was supported by the Australian Research Council. S.V. thanks Nonlinear Physics Center for hospitality and support during his stay in Canberra, and he acknowledges a support of the Serbian Ministry of Science (grant No. OI 141031).

\end{sloppy}
\end{document}